# Unveiling photon-photon coupling induced transparency and absorption

**Kuldeep Kumar Shrivastava[1], Ansuman Sahu[2], Biswanath Bhoi[1,a], Rajeev Singh[1,a]**

[1]Department of Physics, Indian Institute of Technology (Banaras Hindu University), Varanasi - 221005, Bharat (India)

[2]Department of Physical Sciences, Indian Institute of Science Education and Research, Berhampur, Odisha 760003, Bharat (India)

This study presents the theoretical foundations of an analogous electromagnetically induced transparency (EIT) and absorption (EIA) which we are referring as coupling induced transparency (CIT) and absorption (CIA) respectively, along with an exploration of the transition between these phenomena. We provide a concise phenomenological description with analytical expressions for transmission spectra and dispersion elucidating how the interplay of coherent and dissipative interactions in a coupled system results in the emergence of level repulsion and attraction, corresponding to CIT and CIA, respectively. The model is validated through numerical simulations using a hybrid system comprising a split ring resonator (SRR) and electric inductive-capacitive (ELC) resonator in planar geometry. We analyse two cases while keeping ELC parameters constant; one involving a dynamic adjustment of the SRR size with a fixed split gap, and the other entailing a varying gap while maintaining a constant SRR size. Notably, in the first case, the dispersion profile of the transmission signal demonstrates level repulsion, while the second case results in level attraction, effectively showcasing CIT and CIA, respectively. These simulated findings not only align with the theoretical model but also underscore the versatility of our approach. Subsequently, we expand our model to a more general case, demonstrating that a controlled transition from CIT to CIA is achievable by manipulating the dissipation rate of individual modes within the hybrid system, leading to either coherent or dissipative interactions between the modes. The results provide a pathway for designing hybrid systems that can control the group velocity of light, offering potential applications in the fields of optical switching and quantum information technology.

[a]Corresponding author E-mail: rajeevs.phy@iitbhu.ac.in, biswanath.phy@iitbhu.ac.in

# 1. INTRODUCTION

In the last few decades, studies centered around the control and manipulation of electromagnetic (EM) waves have flourished and lead to many interesting physical phenomena with applications in diverse fields ranging from optoelectronics to quantum technology [1-3]. Emerging phenomena stemming from atomic coherence in light-matter interactions are deemed critical for advancing future quantum technology, while the pursuit of precise control over electromagnetic waves continues to inspire extensive research in this domain. Electromagnetically induced transparency (EIT), and its inverse effect absorption (EIA) are the two exotic phenomena of light-matter interactions that provided a foundation for many new and exciting possibilities based on quantum interference effect [4-7]. EIT relies on the coherent interaction between two optical fields, leading to reduced resonant absorption and rendering transparency in a medium or the coupling center, with a wide range of applications including slow light, optical signal processing, quantum switching, four-wave mixing and quantum computation. On the other hand, EIA results from dissipative interaction that enables the absorption of light at coupling center thereby potentiating many applications such as fast light, narrowband filtering, absorption switching, and optical modulators [8-9]. However, EIT and EIA typically owe their existence to non-linear behaviour of the systems. Nonetheless, recent discoveries have identified analogues of EIT and EIA in various systems such as photonic crystals, plasmas, and metasurfaces composed of materials like graphene, vanadium dioxide, and photosensitive silicon [10-14]. Although these analogue effects have garnered significant attention from the scientific communities, there has been comparatively less emphasis on the integration and manipulation of both phenomena within a single device.

More recently, various coupled hybrid systems, such as photon-magnon systems, opto-mechanical systems, and optical microcavity systems, have been proposed to achieve both EIT-like coupling-induced transparency (CIT) and EIA-like coupling-induced absorption (CIA) effects [15-19]. The CIT and CIA effects were observed through the manifestation of mode splitting and merging at or near their coupling center, referred to as level repulsion (LR) and level attraction (LA), respectively [15-19]. Nevertheless, generating CIT or CIA phenomena experimentally often necessitates demanding conditions, including ultra-low temperatures and intricate three-dimensional designs with additional infrastructure, substantially restricting their practical applications [17-19]. Furthermore, in the quest to understand the simultaneous observation of CIT and CIA, different research groups have proposed various theoretical models. Some models are suggesting need of an auxiliary mode [18,20], some suggests use of a hypothetical negative energy mode [18] while the proposed model here effectively explain observed phenomenon [16-23]. Consequently, there is a keen interest among researchers in the development of a miniaturized planar device with a robust theoretical model capable of hosting both types of CIT and CIA quantum phenomena concurrently that would allow a single device to encompass the functionalities of multiple systems. Therefore, the study of CIT/CIA is presently an important area of research, addressing both fundamental principles and practical applications.

In this study, we establish a comprehensive theoretical framework for CIT and CIA, elucidating these phenomena within a microwave photon-photon coupled system. A notable difference from EIT/EIA or several earlier works on CIT/CIA is that our demonstration of the phenomena uses only linear couplings. We demonstrate how a coherent and dissipative coupling between the photon modes leads to the emergence of level repulsion (LR) and attraction (LA), which correspond to CIT and CIA, respectively. Further, we substantiate both LR and LA via numerical simulations using a planar hybrid system comprising two resonating elements that excite photon modes at microwave frequencies. Finally, we achieve a substantial control not only in the coupling strength but also a controllable transition between CIT and CIA by manipulating the dissipation rates. This work presents potential avenues for controlling interactions between two modes, offering a feasible solution for demonstrating CIT and CIA in a single device at room temperature. Such advancements are expected to play a crucial role in the development of simple planar devices for future quantum technology.

## 2. THEORETICAL FORMALISM

To engineer the observation of CIT and CIA for any two interacting modes, we propose a general theoretical model for a hybrid quantum system as schematically shown in Fig. 1. It comprises of two resonating modes A and B, positioned on either side of a common signal line through which both modes are being driven. Both modes, mode A and B are coupled to the signal line and also coupled with each other through a complex coupling constant.

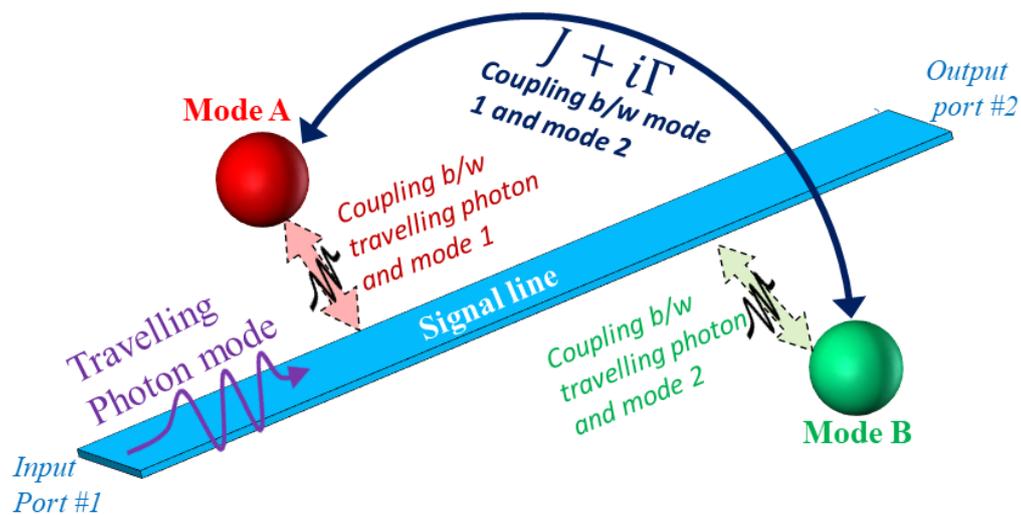

**Fig. 1.** Schematic diagram showing two modes A and B with a common signal line and different attributes of the system.

### 2.1 Hamiltonian for two modes and signal line

The general Hamiltonian for the system involving coherent and dissipative couplings in a two-mode system having common signal line can be written as [21-23]

$$H/\hbar = (\omega_A - i\alpha)\hat{A}^\dagger\hat{A} + (\omega_B - i\beta)\hat{B}^\dagger\hat{B} + \Delta(\hat{A}\hat{B}^\dagger + \hat{A}^\dagger\hat{B}) + \int \omega_k \hat{p}_k^\dagger \hat{p}_k dk$$

$$+ \int [\lambda_A(\hat{A} + \hat{A}^\dagger)(\hat{p}_k + \hat{p}_k^\dagger) + \lambda_B(\hat{B} + \hat{B}^\dagger)(\hat{p}_k + \hat{p}_k^\dagger)]dk \quad (1)$$

Here $\hat{A}$ ($\hat{A}^\dagger$) and $\hat{B}$ ($\hat{B}^\dagger$) are the annihilation (creation) operators of the modes A and B respectively. $\omega_A$ and $\omega_B$ denote the resonance frequencies of the uncoupled modes A and B respectively, while α and β denote the intrinsic damping rates of these two modes respectively, $\Delta = J + i\Gamma$ where J and $\Gamma$ are real parameters that characterize the strength of coherent and dissipative interactions between the two modes.

The fourth term in the Hamiltonian represents the signal line connected to the input and output ports. In our Hamiltonian formulation, we have modelled it as traveling photons, integrating over a real domain from -∞ to +∞. $\hat{p}_k^\dagger$ ($\hat{p}_k$) is bosonic creation (annihilation) operator of the traveling photon obeying $[\hat{p}_k, \hat{p}_{k'}^\dagger] = \delta(k - k')$. The frequency of the travelling photon is denoted by $\omega_k$ where $k$ represents the wave vector. The fifth term accounts for the interaction between each mode and traveling photons, modelled linearly in $\hat{p}_k^\dagger$ and $\hat{p}_k$. Each mode, A and B, exhibits a distinct coupling strength $\lambda_A$ and $\lambda_B$ respectively, arising from the interaction of traveling photons with modes A and B, respectively.

## 2.2 Heisenberg-Langevin Equations

Following the rotating-wave approximation (RWA) and solving Eq. (1), the equation of motion for $\hat{p}_k$ (travelling photon) can be written as [21-25]

$$\dot{\hat{p}}_k = \frac{-i}{\hbar}[\hat{p}_k, H] = -i\omega_k \hat{p}_k - i\lambda_A \hat{A} - i\lambda_B \hat{B} \quad (2)$$

This can be solved to give a time forwarded equation for the travelling wave operator, where the initial time $t_0 < t$ and $\hat{p}_k(t_0)$ is the initial state of the operator

$$\hat{p}_k(t) = e^{-i\omega_k(t-t_0)}\hat{p}_k(t_0) - \int_{t_0}^{t} i[\lambda_A \hat{A} + \lambda_B \hat{B}]e^{-i\omega_k(t-t')}dt' \quad (3)$$

The input field operator at the input port is defined as

$$\hat{p}_{in}(t) = \frac{1}{\sqrt{2\pi}} \int e^{-i\omega_k(t-t_0)} \hat{p}_k(t_0) dk \qquad (4)$$

The equation for mode A operator is

$$\dot{\hat{A}} = \frac{-i}{\hbar}[\hat{A}, H] = -i\tilde{\omega}_A \hat{A} - \int i\lambda_A \hat{p}_k dk - i\Delta\hat{B} \qquad (5)$$

By incorporating Eq. (3) and (4) in Eq. (5), the time forwarded Heisenberg-Langevin equation of the coupled system becomes

$$\dot{\hat{A}}(t) = -i\tilde{\omega}_A \hat{A}(t) - i\sqrt{\gamma}\hat{p}_{in}(t) - \gamma\hat{A}(t) - \sqrt{\kappa\gamma}\hat{B}(t) - i\Delta\hat{B}(t) \qquad (6)$$

Similarly, for the mode B we get

$$\dot{\hat{B}}(t) = -i\tilde{\omega}_B \hat{B}(t) - i\sqrt{\kappa}\hat{p}_{in}(t) - \kappa\hat{B}(t) - \sqrt{\kappa\gamma}\hat{A}(t) - i\Delta\hat{B}(t) \qquad (7)$$

where $\tilde{\omega}_A = \omega_A - i\alpha$ and $\tilde{\omega}_B = \omega_B - i\beta$. $\gamma = 2\pi\lambda_A^2$ and $\kappa = 2\pi\lambda_B^2$ represent the extrinsic damping rates of the mode A and B respectively. The steady state equations in the frequency domain can be obtained by applying a Fourier transformation to Eqs. (6) and (7), yielding [21,22]

$$i(\omega - \tilde{\omega}_A)\hat{A}(\omega) - i\sqrt{\gamma}\hat{p}_{in}(\omega) - \gamma\hat{A}(\omega) - \sqrt{\kappa\gamma}\hat{B}(\omega) - i\Delta\hat{B}(\omega) = 0 \qquad (8)$$

$$i(\omega - \tilde{\omega}_B)\hat{B}(\omega) - i\sqrt{\kappa}\hat{p}_{in}(\omega) - \kappa\hat{B}(\omega) - \sqrt{\kappa\gamma}\hat{A}(\omega) - i\Delta\hat{A}(\omega) = 0 \qquad (9)$$

Further we can write the time retarded equation for the travelling wave operator from Eq. (2), with reference to a later time $t_1 > t$

$$\hat{p}_k(t) = e^{-i\omega_k(t-t_1)}\hat{p}_k(t_1) + \int_t^{t_1} i[\lambda_A \hat{A} + \lambda_B \hat{B}]e^{-i\omega_k(t-t')}dt' \qquad (10)$$

The output field operator at output port is defined as

$$\hat{p}_{out}(t) = \frac{1}{\sqrt{2\pi}} \int e^{-i\omega_k(t-t_1)} \hat{p}_k(t_1) dk \qquad (11)$$

Using Eqs. (10 & 11) in Eq. (5) the time retarded Heisenberg-Langevin equations of the coupled system in the time and frequency domains are

$$\dot{\hat{A}}(t) = -i\tilde{\omega}_A \hat{A}(t) - i\sqrt{\gamma}\hat{p}_{out}(t) + \gamma\hat{A}(t) + \sqrt{\kappa\gamma}\,\hat{B}(t) - i\Delta\hat{B}(t) \tag{12}$$

$$\dot{\hat{B}}(t) = -i\tilde{\omega}_B \hat{B}(t) - i\sqrt{\kappa}\hat{p}_{out}(t) + \kappa\hat{B}(t) + \sqrt{\kappa\gamma}\,\hat{A}(t) - i\Delta\hat{A}(t) \tag{13}$$

$$i(\omega - \tilde{\omega}_A)\hat{A}(\omega) - i\sqrt{\gamma}\hat{p}_{out}(\omega) + \gamma\hat{A}(\omega) + \sqrt{\kappa\gamma}\,\hat{B}(\omega) - i\Delta\hat{B}(\omega) = 0 \tag{14}$$

$$i(\omega - \tilde{\omega}_B)\hat{B}(\omega) - i\sqrt{\kappa}\hat{p}_{out}(\omega) + \kappa\hat{B}(\omega) + \sqrt{\kappa\gamma}\,\hat{A}(\omega) - i\Delta\hat{A}(\omega) = 0 \tag{15}$$

**2.3 Transmission**

Either solving Eqs. (8 and 14) or Eqs. (9 and 15) as we can get the relation between input and output fields as

$$\hat{p}_{out} = \hat{p}_{in} - 2i\sqrt{\kappa}\,\hat{B}(\omega) - 2i\sqrt{\gamma}\,\hat{A}(\omega) \tag{16}$$

We can solve Eqs. (8 and 9) to get the field operators in terms of the input field as

$$\hat{A}(\omega) = \frac{\hat{p}_{in}(\Delta\sqrt{\kappa} + i\beta\sqrt{\gamma} + \sqrt{\gamma}\omega - \sqrt{\gamma}\omega_B)}{(i\Delta + \sqrt{\kappa\gamma})^2 + (i\alpha + i\gamma + \omega - \omega_A)(i\beta + i\kappa + \omega - \omega_B)} \tag{17}$$

$$\hat{B}(\omega) = \frac{\hat{p}_{in}(\Delta\sqrt{\gamma} + i\alpha\sqrt{\kappa} + \sqrt{\kappa}\omega - \sqrt{\kappa}\omega_A)}{(i\Delta + \sqrt{\kappa\gamma})^2 + (i\alpha + i\gamma + \omega - \omega_A)(i\beta + i\kappa + \omega - \omega_B)} \tag{18}$$

The experimentally measurable and numerically simulated parameter for these kinds of experiments is the transmission coefficient $S_{21}$ between ports 1 and 2 of the hybrid system, which for our geometry gives

$$S_{21} = \frac{\hat{p}_{out}}{\hat{p}_{in}} - 1 \tag{19}$$

Utilizing Eqs. (16, 17, and 18) in Eq. (19) yields the $S_{21}$ as follows

$$S_{21} = \frac{2\alpha\kappa + 2\beta\gamma - 4i\Delta\sqrt{\kappa\gamma} - 2i\gamma(\omega - \omega_B) - 2i\kappa(\omega - \omega_A)}{(i\Delta + \sqrt{\kappa\gamma})^2 + (i\alpha + i\gamma + \omega - \omega_A)(i\beta + i\kappa + \omega - \omega_B)} \tag{20}$$

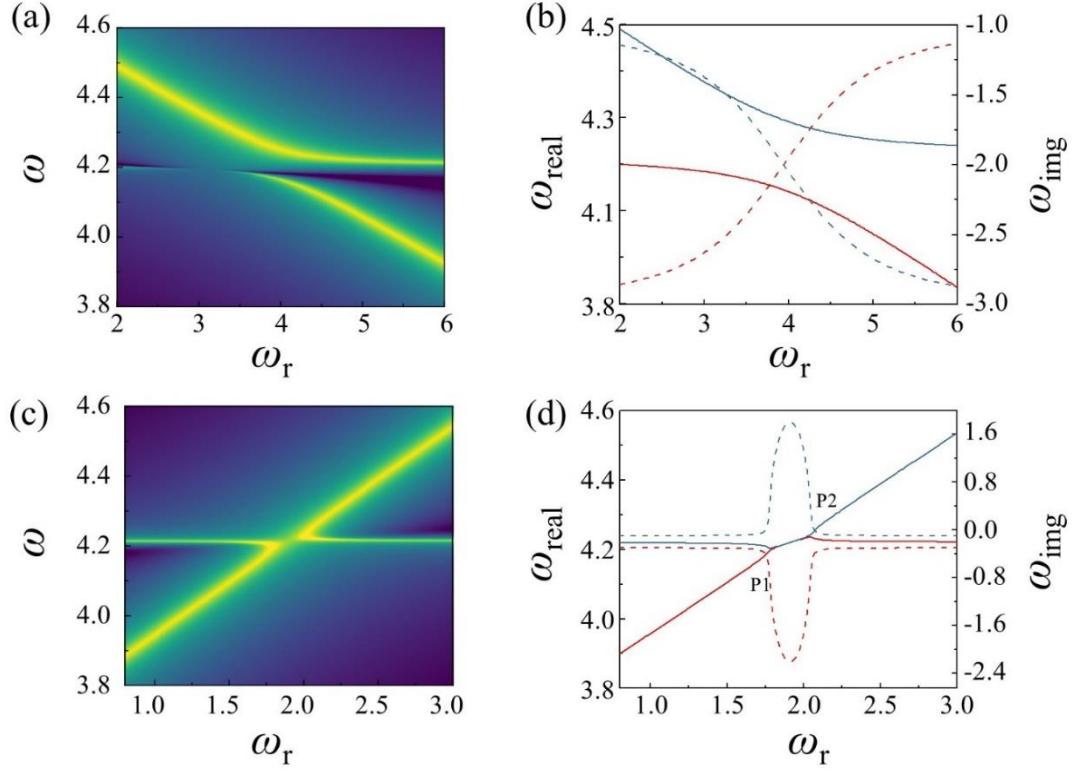

**Fig. 2.** Numerical calculations of dispersion spectra (a) for case-1 (J > Γ) and (c) for case-2 (Γ >J). Numerical calculations of resonance frequencies (solid lines) and linewidths (dotted lines) (b) for case-1 (J > Γ) and (d) for case-2 (Γ >J).

We will begin by examining the conditions under which the CIT and CIA phenomena manifest, each of which will be exemplified by the notions of LR and LA, respectively. We use extrinsic damping constants for the mode A, $\gamma = 0.01$, and mode B, $\kappa = 0.001, 0.0001$ which are consistent with experimental values of hybrid systems [15,18,20,22,23]. The resonance frequencies of mode A and B are $\omega_A = 4.22$ and $\omega_B = 3.84 \; to \; 4.59$ GHz with intrinsic damping $\alpha = 0.001$, and $\beta = 0.001$ respectively [15,18,20,22,23]. The coupling constant is $J + i\Gamma'$, $where \; \Gamma' = \Gamma + \sqrt{\gamma\kappa}$. Using Eq. (20), we numerically calculate the |S$_{21}$| power on the $\omega_r - \omega$ plane for two different the conditions; (i) $J \gg \Gamma'$ and (ii) $\Gamma' \gg J$. Under the dominance of J (i.e. J ≫ Γ'), a normal anticrossing dispersion pattern (CIT) emerges, revealing two higher and lower coupled branches around the resonant frequencies of modes A and B, indicative of the robust coherent coupling between them [see Fig. 2(a)]. On the other hand, for case of dominating $\Gamma'$ (i.e. $\Gamma' \gg J$), a contrasting shape of anticrossing dispersion (i.e., CIA) is observed, as shown in Fig. 2(c). Quite recently, such opposite anticrossing dispersion was also demonstrated, but in complex hybrid systems at a very low temperature below 50 mK [17,19].

To understand the similarity and difference of such CIT and CIA dispersion shapes that depends on the coherent and dissipative interactions between the modes A and B, a coupling matrix is constructed as given by,

$$H_{coupling} = \begin{pmatrix} \widetilde{\omega}_{A'} & \Delta' \\ \Delta' & \widetilde{\omega}_{B'} \end{pmatrix} \quad (21)$$

where $\widetilde{\omega}_{A'} = \omega_A - i\alpha'$ and $\widetilde{\omega}_{B'} = \omega_B - i\beta'$ with $\alpha' = (\alpha + \gamma)$ and $\beta' = (\beta + \kappa)$. The effective coupling constant $\Delta' = J + i\Gamma'$ with $\Gamma' = \Gamma + \sqrt{\gamma\kappa}$. Eq. (21) can be written as

$$H_{coupling} = \frac{\widetilde{\omega}_{A'} + \widetilde{\omega}_{B'}}{2}\mathbb{I} + \frac{\omega_A - \omega_B}{2}\sigma_z + J\sigma_x - i\frac{\alpha' - \beta'}{2}\sigma_z + i\Gamma'\sigma_x \quad (22)$$

where $\mathbb{I}$ is identity matrix and $\sigma_x, \sigma_z$ are Pauli matrices. The term proportional to the identity just causes a shift in the complex eigenvalues. In a system with a Hermitian coupling matrix, normal anticrossing or level repulsion (LR) is observed, while in the case of a purely anti-Hermitian matrix, opposite anticrossing or level attraction (LA) occurs [18,26]. So, we can express the coupling matrix in the general form $C\sigma_c + iD\sigma_d$, with real coefficients C and D. In this general form, level repulsion (attraction) occurs when the Pauli matrix with a dominating real (imaginary) coefficient is present [18,26]. When other parameters are held constant, a dominant real coefficient, particularly when J>>$\Gamma'$, leads to LR, whereas reversing this condition results in LA.

From the eigenvalue analysis of the coupling matrix, we get higher (lower) branches of eigenfrequency given by

$$E_\pm = \frac{\tilde{\omega}_{A'} + \tilde{\omega}_{B'} \pm \sqrt{4(\Delta')^2 + (\tilde{\omega}_{A'} - \tilde{\omega}_{B'})^2}}{2} \qquad (23)$$

Using this equation, we have numerically calculated the complex eigenvalues of two coupled modes, i.e., $E_\pm = \omega_{real} \pm i\omega_{img}$, where $\omega_{real}$ and $\omega_{img}$ represent the dispersion shape and the linewidth evolution of the coupled modes, respectively.

For two specific cases of $J \gg \Gamma'$ and $\Gamma' \gg J$, the resultant numerical calculations are given in Figs. 2(b) and 2(d), respectively. For the case of dominating $J$, the dispersion shape (real value $\omega_{real}$) of $E_+$ and $E_-$ branches repel each other [Fig. 2(b)], while their linewidths (imaginary value $\omega_{img}$) of $E_+$ and $E_-$ cross each other [Fig. 2(b)]. This behaviour is ubiquitous in any coupled systems involving reciprocal energy transfers as reported in Refs. [15-19]. It has been observed in a multimode optomechanical circuit with dissipation engineering [18], between magnon mode of two YIG spheres mediated by cavity photon of a superconducting resonators [19]. On the other hand, for the case of dominating $\Gamma'$, the dispersion shape ($\omega_{real}$) of $E_+$ and $E_-$ provide the energies of their states that attract each other and nearly meet at two points [denoted as P1 and P2 in Fig. 2(d)], while their linewidth (imaginary value $\omega_{img}$) of $E_+$ and $E_-$ are found to be repulsive (they do not cross each other), as shown in the Fig. 2(d). Such LA has been studied in a 3D microwave cavity using an YIG sphere [21], between magnon mode of two YIG spheres mediated by propagating photons of a coplanar waveguide [19].

## 3. RESONATOR DESIGN AND NUMERICAL SIMULATION SET UP

To gain a detailed quantitative insight into the theoretical model, we employed the electromagnetic full-wave solver CST Microwave Studio to design and conduct numerical simulations on a planar hybrid system comprising two photon resonators. Among the two resonators, one is a split-ring resonator (SRR) while the other is an electric inductive-capacitive resonator (ELC), as schematically shown in Fig. 3(a) and Fig. 3 (d), respectively. In the microstrip line configuration, electric fields emanate from the central strip, terminating perpendicular to the ground plane, while a microwave magnetic field form around the microstrip line. The resonators were positioned in close proximity to the microstrip line, allowing them to be excited by the induced currents from the transverse microwave magnetic field when microwave currents flow through the microstrip feeding line. In this state, the resonators function as a parallel LC resonant circuit, resulting in a quasi-static resonant effect.

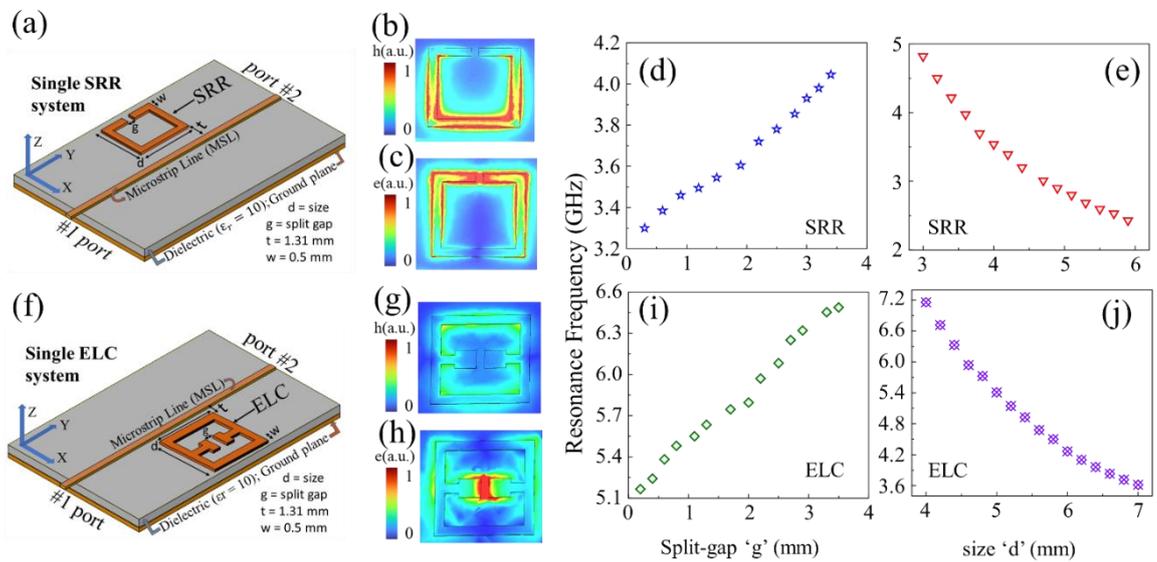

**Fig. 3.** Schematic drawing planar resonators loaded microstripline for numerical simulation (a) split ring resonator (SRR) and (f) electric inductive-capacitive resonator (ELC). Electric and magnetic fields distributions for SRR and ELC are shown in (b, c) and (g, h) respectively. Dependence of resonance frequency on split gap size (d, i) and resonator size (e, j) for (d, e) SRR and (i, j) ELC.

### 3.1 Numerical simulation of individual SRR and ELC

Before considering the hybrid system, we investigate the fine tunability in the resonance frequency for both the SRR and ELC by adjusting their dimensions. The microstrip line was designed to have a width $w = 0.64$ mm, a length $l = 40$ mm, and a thickness $t = 35$ μm in order to have a 50 Ω characteristic impedance. The planar substrate (Gray colour) had dimensions of 50 mm, 25 mm, and 0.64 mm, a relative permittivity of 10, and a dissipation factor of 0.0012 at 10 GHz. The SRR is configured with dimensions of $d = 5$ mm, and $g = 0.4$ mm, as illustrated in Fig. 3(a). By keeping the other geometrical parameters of the SRR constant, we conduct numerical simulation systematically varying the split gap '$g$' from 0.3 to 3.4 mm. This results in an increase in the resonance frequency from 3.3 GHz to 4.05 GHz as '$g$' varies between 0.3 and 3.4 mm as shown in Fig. 3 (b). On the other hand, the resonance frequency decreases from 4.82 GHz to 2.43 GHz as the size of the SRR changes from 3 to 5.9 mm for a fixed value of '$g$' = 0.4 mm (Fig. 3 (c)). Similarly, a separate simulation is performed for the ELC with geometrical parameters of $d = 5$ mm, and $g = 0.2$ to 3.5 mm, as indicated in Fig. 3 (e). Like SRR, its resonance frequency increases from 5.2 GHz to 6.5 GHz as '$g$' is varied between 0.2 and 3.5 mm (for fixed $d = 5$mm), but decreases from 7.2 GHz to 3.6 GHz as '$d$' changes from 4 to 7 mm (for fixed $g = 0.4$ mm) as shown in Fig. 3(e) and 3(f) respectively. These variation in resonance frequency with geometrical parameters can be well understood by viewing the resonators as basic $LC$ circuits, with a resonance frequency $\omega/2\pi = 1/2\pi\sqrt{LC}$, where the inductance (L) and capacitance (C) arise from the conducting path and the split gaps of the resonators respectively. As the split gap widens, the capacitance ($C = \varepsilon A/g$) decreases, leading to an increase in the resonance frequency of the resonators. Conversely, increasing the dimensions of the resonators extends the conducting path, resulting in a decrease in the resonance frequency of the resonators. This relationship between the geometrical parameters and the resonant response provides a guide to design the compact planar photon-photon hybrid system based on the transmission lines loaded with SRR and ELC.

## 3.2 Numerical simulation of SRR/ELC hybrid system

In order to understand and uncover the intricate dynamic interactions between two photon modes, we design SRR and ELC in a single hybrid system by closely placing them on either side of the microstripline as shown in Fig. 4 (a). When an ac current of microwave frequency is transmitted through the microstripline, a substantial portion of the microwave magnetic field lines generated by the microstrip is anticipated to permeate both the SRR and ELC, resulting in a strong electrodynamic coupling between them. This interaction between the photon modes of the SRR and ELC facilitates an energy exchange through their overlapping electric and magnetic fields. This leads to the hybridization of resonance modes, which can be observed in phenomena such as mode splitting by analysing the intensity distribution.

The changes in resonators' electric and magnetic fields, driven by factors like geometry, orientation, position, and electromagnetic properties, impact the constructive or destructive interference that affects the electromagnetic coupling of SRR and ELC. Therefore, the simulation is conducted for two cases: For case-1 the SRR split gap is fixed (at $g = 0.6$ mm), while SRR size '$d$' varies from $2 \times 2$ mm$^2$ to $6 \times 6$ mm$^2$. For case-2, SRR size is fixed ($d = 4 \times 4$ mm$^2$) while the split gap '$g$' varies from 0.8 to 3.2 mm. The other dimensions such as outer dimension ($5 \times 5$ mm$^2$), and split gap (0.2 mm) and size of ELC resonator is fixed for all cases.

## 4. COUPLING BETWEEN SRR's AND ELC's PHOTON MODES

We record the transmission parameter ($S_{21}$) of the coupled SRR-ELC hybrid system as a function of the microwave frequency $f$ of the ac current flowing in the microstrip line for different values of SRR's size and SRR's split gap representing case-1 and case-2 respectively. Figure 4 (b) shows the $|S_{21}|$ spectra of SRR-ELC hybrid system under simulation conditions corresponding to case -1, revealing the presence of two distinct peaks. One peak (marked by red arrows) is very strongly dependent on the SRR's size and continuously shifts towards the lower-frequency side with increasing unit size, and crosses the other peak position (blue arrows), thus indicating the SRR's photon mode. The other peak does not move much, thus indicating the ELC mode. It is worth noting that as one peak approaches the other, there is a gradual change in their amplitudes. The maximum change occurs precisely as one peak crosses over the other, after which they gradually return to their original magnitudes.

Similarly, the interaction between SRR and ELC for case-2 is shown in Fig. 4 (c). As the SRR split gap increases, the SRR mode approaches the ELC mode and combines into a unified mode at the coupling center, resulting in an increase in amplitude. Continued widening of the SRR split gap causes the unified mode to separate into two distinct peaks of SRR and ELC. The shifts in resonance frequency positions and amplitudes depicted in Fig. 4(b) and (c) demonstrate a robust electrodynamics interaction between the SRR and ELC photon modes in the hybrid system. However, a noteworthy observation is that as the SRR size increases (case-1), the peaks approach each other but never coalesce into a single peak. Instead, they exhibit a transparency region at the coupling center, indicating a distinct level repulsion or CIT phenomenon. On the other hand, increase in SRR split gap (case-2), resulted a merging of two modes into a single mode indicating level attraction or CIA phenomenon.

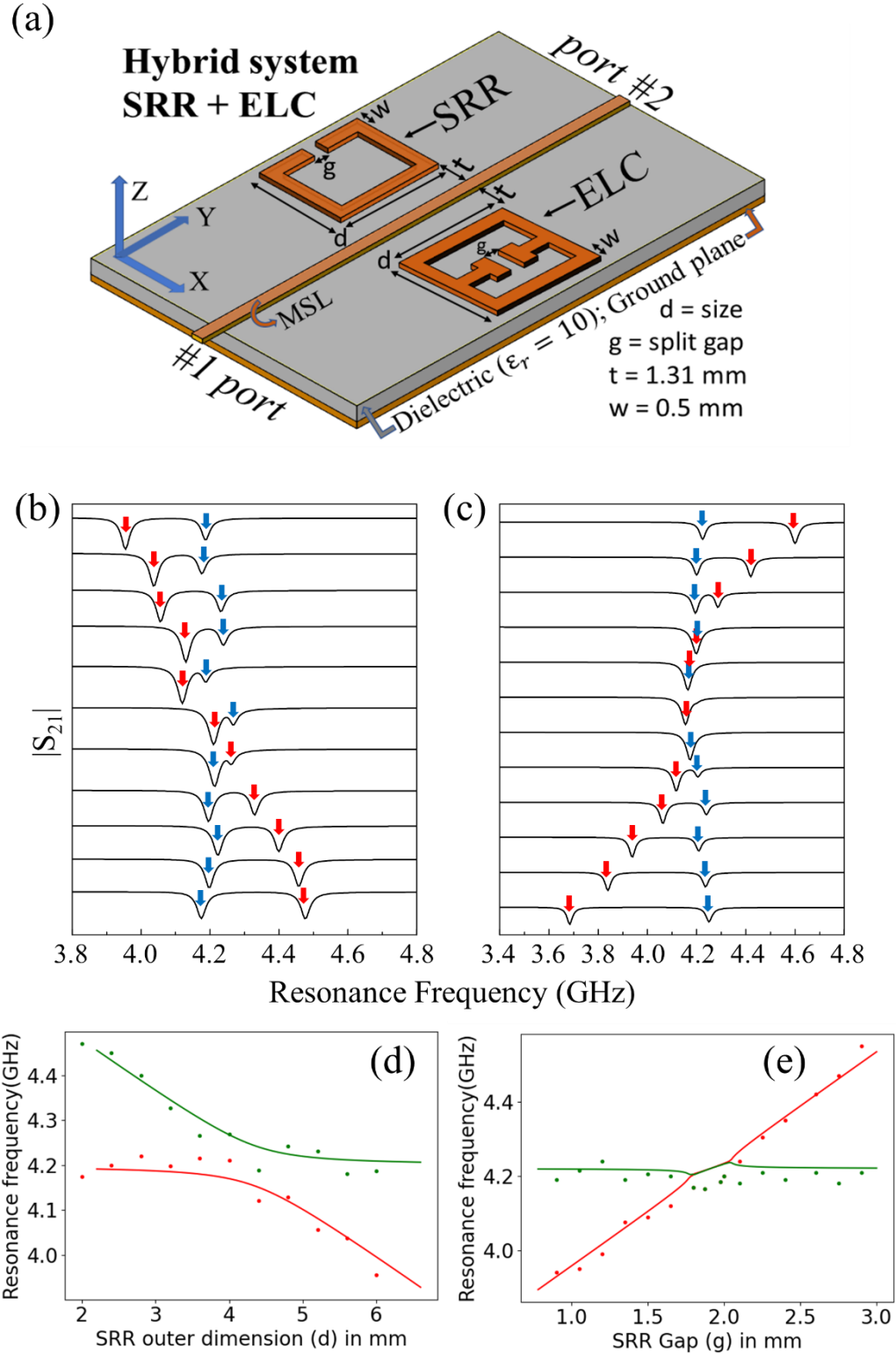

**Fig. 4.** (a) Schematic drawing of planar hybrid system comprising of SRR and ELC. |S$_{21}$| spectra of SRR-ELC hybrid sample as function of (b) SRR size and (c) SRR split gap. Frequencies of the resonance peaks from (b, c) as functions of (d) SRR size and (e) SRR split gap. Dots: simulation data; solid lines: Numerically fitted with Eq. 5.

In order to examine the coupling effect, we extract the resonance frequency values from the |S$_{21}$| spectra and plot a graph illustrating the relationship between the SRR geometrical parameter and the corresponding hybrid resonance frequencies in Fig. 4 (d, e). This graphical representation allows us to observe how changes in the SRR size/split gap influence the resonant behaviour of the coupled SRR-ELC system. In order to understand such contrasting anticrossing effects observed for the case-1 and case-2 geometries, we estimate the coupling constant by fitting with Eq. (23) to the lower- and higher-frequency branches (solid lines) of the coupled modes. From the fitting we obtained the numerical values of coupling constant $J = 0.075\ and\ \Gamma' = 0.02$ MHz for the CIT [Fig. 4(d)] and the CIA [Fig. 4(e)], respectively. The simulated results are well fitted with Eq. (23) obtained from our theoretical modelling. We note that for the CIA, we obtain only the imaginary value of $\Delta'$, whereas for the CIT, the real value of $\Delta'$. Therefore, the real and imaginary values for $\Delta'$ characterize the CIT and the CIA phenomena, respectively. According to Eq. (23), the real and imaginary numbers of $\Delta'$ are the cases of J dominating and $\Gamma'$ dominating respectively.

## 5. TRANSITION BETWEEN CIT AND CIA

In order to understand the effect of various controlling parameters determining the contrasting dispersions during photon-photon interactions and the transitions between CIT and CIA, we further analyse the general coupling matrix with all dissipation and coupling terms present. From Eq. (23) we can write

$$E_+ - E_- = \sqrt{4(\Delta')^2 + (\widetilde{\omega}_{A'} - \widetilde{\omega}_{B'})^2} \qquad (24)$$

For level attraction the real (imaginary) part of eigenfrequencies should (should not) cross each other. In other words, the real part of $E_+ - E_-$ is zero for some input frequency, while its imaginary part will never be zero. At the crossing of the real part, we get $(E_+ - E_-)^2_{imaginary\ part} = 0$ and $(E_+ - E_-)^2_{real\ part} < 0$. Solving together gives,

$$|J| < \frac{|\alpha' - \beta'|}{2} \qquad (25)$$

$$\Gamma' = \frac{(\alpha' - \beta')(\omega_A - \omega_B)}{4J} \qquad (26)$$

Thus, to observe level attraction, the coherent coupling (J) should be small while the dissipative coupling ($\Gamma'$) should be large. To explore the effect of damping of each mode and how that is dominating the level attraction and level repulsion phenomenon, and how their interplay is affected by other parameters we have plotted 3D colour map for the different range of parameters. Our final result depends on four variables $\alpha', \beta', J$ and $\Gamma'$ apart from the resonators eigenfrequencies which we keep fixed. We consider two specific cases (case-1 and case-2) for varying $J$ and $\Gamma'$.

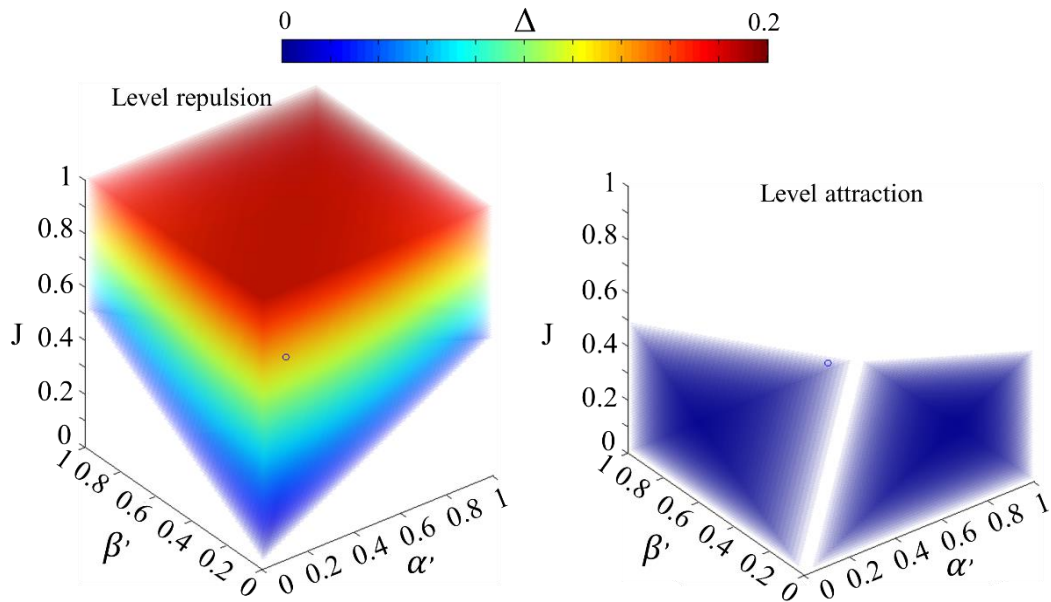

**Fig. 5.** Phase diagram of CIT and CIA for total dissipation for both the modes and J while keeping T=0. (a) level repulsion i.e. CIT and (b) level attraction i.e. CIA.

Fig. 5 represents a pictorial overview of different possibilities of the coupling Hamiltonian in terms of $\alpha'$, $\beta'$ and $J$ or $\Gamma'$ as variable. For case-1 Fig. 5(a) and (b) shows the 3D colourmap where we have taken, $\Gamma' = 0.001$, and have plotted the real part of the difference of the eigenvalues for different values of $\alpha'$, $\beta'$ and $J$. Fig. 5 (a) shows the region of level repulsion which violates Eq. (25). We clearly see that J is dominating for the level repulsion Fig. 5 (b) shows the region of level attraction which is complementary to that of level repulsion. We observe from Fig. 5(b) that for $\Gamma' = 0.001$, there is a large region of level attraction at J=0 which decreases gradually as J is increased. This region is symmetric about the $\alpha' = \beta'$ line. This decrease continues till J=0.05 above which there is only level repulsion. The parameters of the hybrid system we considered for case-1 is a point ($J = 0.075$, $\alpha' = 0.011$ and $\beta' = 0.0011$) shown in the 3D colourmaps which is inside the Fig.5(a) but outside the Fig.5(b). As mentioned before the two figures Fig.5(a) and Fig.5(b) are complementary to each other and their boundary have may also have exceptional points which have gained significant interest lately in exploring the topological nature of coupled hybrid systems [18,27].

## 6. CONCLUSION

We establish a comprehensive theoretical framework for observation of CIT and CIA in a coupled hybrid system, which is verified through numerical simulation for photon-photon interaction in a planar SRR-ELC hybrid system. The dominant coherent and dissipative interaction between the photon modes leads to CIT and CIA respectively. Theoretical calculations, in excellent agreement with simulations, reveal the physical origins of various interactions, enabling substantial control over both coupling strength and a controllable transition between CIT and CIA through manipulation of the dissipation rates of the two modes. Our theoretical model combined with numerical calculations, can contribute to understanding the physics of coupled systems and investigating interactions among excitations such as plasmons, magnons, and phonons, which are valuable for quantum information processing. Moreover, this work can open several theoretical and experimental endeavors to explore similar phenomena that potentially help in the advancement of concepts important for the development of quantum photonic and nonreciprocal quantum devices.

## Acknowledgements

The work was supported by the Council of Science & Technology, Uttar Pradesh (CSTUP), (Project Id: 2470, CST, U.P. sanction No: CST/D-1520). B. Bhoi acknowledges support by the Science and Engineering Research Board (SERB) India- SRG/2023/001355

## Statements & Declarations

**Funding:** The authors declare that grants were received from CSTUP and SERB during the preparation of this manuscript.

**Competing Interests:** The authors declare that they have no competing interests.

**Author Contributions:** All authors contributed to the study conception and design. R. S and B. B led the work and wrote the manuscript with K.S. The other co-authors read, commented and approved the final manuscript.

## Data Availability

The data that support the findings of this study are available within the article.